\documentstyle[12pt]{article}

\def\be{\begin{equation}}
\def\ee{\end{equation}}
\def\bq{\begin{eqnarray}}
\def\eq{\end{eqnarray}}
\def\ra{\rightarrow}
\def\n{\nonumber}
\def\g{\gamma}
\def\vp{\varphi}

\begin{document}

\begin{flushright}
SPhT-t97/032\\
hep-ph/9703244
\end{flushright}
\vspace{1cm}
\begin{center}
{\bf Twist-2 Light-Cone Pion Wave Function. }
\end{center}
\begin{center}{V.M. Belyaev$^*$ }
\end{center}
 \begin{center}
 {\em SPhT, CEA-SACLAY, 91191 Gif-sur-Yvette, CEDEX, France}
\end{center}
\begin{center}{Mikkel B. Johnson}  \\
{Los Alamos National Laboratory, Los Alamos, NM 87545, USA
}
\end{center}
\vspace{1cm}
\begin{abstract}
We present an analysis of the existing constraints for the twist-2
light-cone pion wave function.
We find that existing information on the pion wave function
does not exclude the possibility that the pion wave function
attains its asymptotic form.
New bounds on the parameters of the pion wave function are
presented.

\vspace{0.5cm}

\noindent PACS number(s): 11.15.Tk, 11.55.Hx,  12.38.Lg, 13.60.Hb
\end{abstract}

\vspace{1cm}
\flushbottom{$^*\overline{On\;  leave\;  of\;  absence }
\;from\;  ITEP,\;  117259\;  Moscow,\; Russia.$}

 \newpage

The light-cone wave function was introduced in perturbative 
quantum chromodynamics (QCD) to
describe hadron form factors at large $Q^2$ \cite{cz,er,bl}.  At the present 
time, the pion light-cone wave function $\vp_\pi(u)$ is widely 
used in perturbative and nonperturbative 
(in the framework of QCD sum rule approach \cite{SVZ}) descriptions of hadron properties.
For example, numerous hadron amplitudes \cite{light} have been determined
from light-cone QCD sum rules suggested in \cite{bbk}.
In this approach,  the pion light-cone wave function $\vp_\pi(u)$ is given as the following matrix
element of a twist-two operator, 
\bq
<0|\bar{u}(x)\g_\mu\g_5d(0)|\pi(q)>_{x^2=0}=if_\pi 
q_\mu\int_0^1e^{-iu(qx)}\vp_\pi(u)du.
\label{vp}
\eq
As results depend on the model of light-cone wave functions, it is clearly
important to understand which forms for the wave functions 
are consistent with the existing constraints.  In this paper, we analyze
all such constraints on $\vp_\pi(u)$, including the new bounds determined 
from a recent analysis of the structure function of the pion.

There are three relevant models for $\vp_\pi(u)$.  These are the asymptotic 
wave function,
\bq
\vp_\pi^{as.}(u)=6u(1-u),
\label{1}
\eq
the Chernyak-Zhitnitsky wave 
function $\vp_\pi^{CZ}(u)$,
\bq
\vp_\pi^{CZ}(u)=30 u(1-u)(2 u-1)^2,
\label{2}
\eq
which was suggested to describe the pion form factor for $Q^2\sim 5-10GeV^2$ 
in perturbative QCD, and the wave function of Braun and Filyanov 
$\vp_\pi^{BF}(u)$,
\bq
\vp_\pi^{BF}(u)=6u(1-u)\{1+a_2(\mu) 3/2[5(2 u-1)^2-1]\n
\\
+a_4(\mu) 15/8[21(2 u-1)^4-14(2 u-1)^2+1]\}
\n
\\
(a_2\simeq 0.44,\;\; a_4\simeq 0.25;\;\;\;\mu\simeq 1GeV),
\label{3}
\eq
where $\mu$ is the normalization point.  
The model of Braun and Filyanov was obtained from the light-cone QCD sum rule 
for the coupling constant $g_{\pi NN}$ \cite{bk,bf}  giving the following constraint
on the light-cone wave function:
\bq
\vp_\pi(0.5)=1.2.
\label{4a}
\eq
The value of the second moment was obtained in \cite{cz}:
\bq
 m_2=\int_0^1 u^2\vp_\pi(u)du=0.35;\;\;\;\mu\simeq 1GeV.
\label{4b}
\eq
However it was pointed that the  QCD sum rule for the second
moment of the pion wave function
 is not accurate enough to decide in favor of the 
Chernyak-Zhitnitsky wave function or the asymptotic one (see
for example \cite{mnpls}).
The asymptotic value $m_2=0.3$ is not excluded.
So, it means that  $m_2$ is know with accuracy not
better that $15\%$. This accuracy we will use in our analysis.

There is an additional constraint formulated by Radyushkin and Rustkov \cite{rad}:
\bq
I=\int_0^1\frac{\vp_\pi(u)}{u}du=2.4
\label{5}
\eq
This relation was obtained from the QCD sum rule for the transition form factor
$\g\g^*\ra\pi^0$, which was compared with prediction of perturbative QCD. 
The authors compare their result (\ref{5}) with the different models,

\bq
I^{as.}=\int_0^1\frac{\vp^{as.}_\pi(u)}{u}du=3
\label{6}
\\
I^{CZ}=\int_0^1\frac{\vp^{CZ}_\pi(u)}{u}du=5
\label{9}
\\
I^{BF}=\int_0^1\frac{\vp^{BF}_\pi(u)}{u}du=5.07,
\label{10}
\eq
and interpret this as an indication
that the pion wave function is not very different from its asymptotic form.
Unfortunately, the authors in \cite{rad} did not
discuss the accuracy of their results and we can not use this 
result (\ref{5}) in our analysis.

Finally, we use the new constraint  for $\vp_\pi(u)$ obtained from an analysis 
of the light-cone 
QCD sum rule for the pion structure function \cite{bj, bj1}, which gives
\bq
\vp_\pi(0.3)=1;\;\;\; \mu\simeq 1 GeV.
\label{11}
\eq

In what follows, we will consider constraints for $m_2$, $\vp_\pi(0.5)$ 
and $\vp_\pi(0.3)$.
The accuracy of the constraints for $\vp_\pi(0.5)$ and $\vp_\pi(0.3)$ is
about 20\%. 

To begin our analysis of the existing constraints for the pion wave function,
we have to choose a reasonable parametrization.  We use the results of 
\cite{bf1}, in which a series expansion of light-cone wave functions was 
suggested with the higher-order terms corresponding to operators with 
increasing conformal spin.
In the case of the twist-2 pion wave function, this expansion is
\bq
\vp_\pi(u)=6 u(1-u)\left\{ 1+a_2 C_2^{3/2}(2u-1)+a_4 C_4^{3/2}(2u-1)
\right.
\n
\\
\left.
+a_6 C_6^{3/2}(2u-1)+...\right\}.
\label{12}
\eq
Here $C_n^{3/2}$ are the Gegenbauer polynomials; $C_2^{3/2}(x)=\frac32(5x^2-1)$,
$C_4^{3/2}(x)=15/8(21x^4-14x^2+1)$.
If we assume that the pion wave function is not very different from its
asymptotic form, then we can expect that the higher terms  in (\ref{12})
are small. 
This assumption means that there should be the following
relations:
\bq
1\gg a_2\gg a_4 \gg a_6 \gg ...
\label{13}
\eq
In the present analysis we take into consideration only the three 
leading terms in the expansion
(\ref{12}): $1,a_2,a_4$.  At the end of our analysis we demonstrate the 
presented constraints indicate in favor of
validity of the relations (\ref{13}).

The constraint (\ref{4a}) in the parametrization (\ref{12}) has the following 
form,
\bq
m_2=\int_0^1 u^2\vp_\pi(u)du=\frac{3}{70}(7+2 a_2)=0.35\pm 0.05.
\label{14}
\eq
Note that the higher  terms of expansion (\ref{12}) do not contribute to the
relation (\ref{14}).
The second constraint (\ref{4b}) leads to the following result,
\bq
\vp_\pi(0.5)=\frac32\left( 1-\frac32 a_2+ \frac{15}8 a_4\right)=1.25\pm 0.25.
\label{15}
\eq
Using relation (\ref{5}) in the parametrization (\ref{12}) gives 
\bq
I=\int_0^1\frac{\vp_\pi(u)}{u}du=3(1+a_2+a_4)=2.4.
\label{16}
\eq
And, the last constraint (\ref{11}) gives us the following  formula:
\bq
\vp_\pi(0.3)=1.26 (1 - 0.3 a _2- 1.317 a_4)=1\pm 0.2.
\label{17}
\eq

It is convenient to present all existing constraints on a plot with 
axis $a_2,a_4$ (see Fig.1).
Note that the hatched region in Fig.1 is in agreement with our
assumption on the hierarchy (\ref{13}).
Points corresponding to the asymptotic wave function ($a_2=a_4=0$),
the Chernyak-Zhitnitsky wave function ($a_2=\frac23,a_4=0$),
and the Braun-Filyanov wave function ($a_2=0.44,a_4=0.25$)
are also shown on this plot.

Note that due to the relatively small coefficient of $a_2$ in eq.(\ref{14}),
a small uncertainty in the value of $m_2$ leads to a big uncertainty 
for $a_2$.  Assuming that $m_2=0.35\pm 0.05$ we obtain 
\bq
0<a_2<1.2.
\label{18}
\eq

The relation (\ref{18}) does not determine the value of $a_2$ very accurately,
but it is useful, showing that $a_2>0$.

The constraints for $\vp_\pi(0.5)$ and $\vp_\pi(0.3)$ are more sensitive to the
parameters $a_2$ and $a_4$.
From relations (\ref{15},\ref{17}) it follows
that
\bq
a_2=0.25\pm 0.25;\;\;\; a_4=0.1\pm 0.12,
\label{19}
\eq
and we can not exclude that the pion wave function attains its asymptotic form.
From relation (\ref{19}) we obtain the following prediction:
\bq
I=4\pm 1.
\label{24}
\eq

Note that the best representation of the 
quark distribution \cite{bj,bj1} was obtained from the light-cone QCD 
sum rule for the case when
the pion wave function is very close to its asymptotic form.
This can be used as an argument in the favor of suggestion
that pion wave function is nearly asymptotic.

In summary, we have presented an analysis of the known constraints for the
twist-2 pion light-cone wave function. We have accordingly found new bounds 
on the form of the pion wave function. 
We note that the light-cone QCD sum rule for the quark distribution in a pion
indicates that $\vp_\pi(u)$ is close to its asymptotic form.

This research was sponsored in part by the U.S. Department of Energy
at Los Alamos National Laboratory under contract 
W-7405-ENG-36.

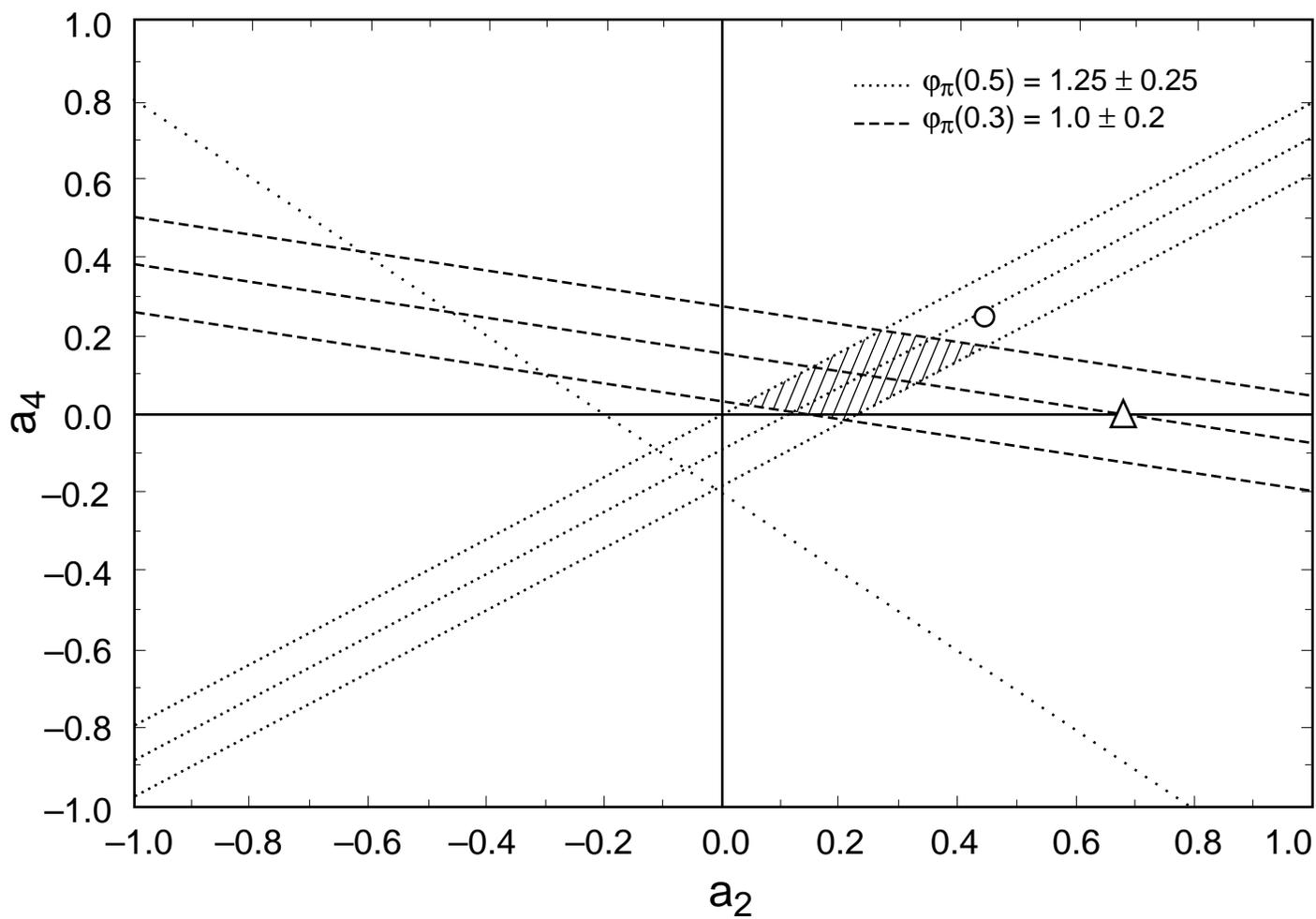
\end{document}